\documentclass[10pt, paper=a4, UKenglish]{article}
\usepackage{graphicx}

\def\Title#1{\begin{center} {\Large #1 } \end{center}}
\def\Author#1{\begin{center}{ \sc #1} \end{center}}
\def\Address#1{\begin{center}{ \it #1} \end{center}}

\newcommand\pubblock{\rightline{\begin{tabular}{l} Proceedings of CTD 2020\\ \pubnumber\\
         \pubdate  \end{tabular}}}

\newenvironment{Abstract}{\begin{quotation} \begin{center} 
             \large ABSTRACT \end{center}\bigskip 
      \begin{center}\begin{large}}{\end{large}\end{center} \end{quotation}}

\newenvironment{Presented}{\begin{quotation} \begin{center} 
             PRESENTED AT\end{center}\bigskip 
      \begin{center}\begin{large}}{\end{large}\end{center} \end{quotation}}

\def\beq{\begin{equation}}
\def\eeq#1{\label{#1}\end{equation}}
\def\eeqn{\end{equation}}

\def\beqa{\begin{eqnarray}}
\def\eeqa#1{\label{#1}\end{eqnarray}}
\def\eeqan{\end{eqnarray}}

\let\bar=\overbar

\def\Dslash{\not{\hbox{\kern-4pt $D$}}}
\def\dslash{\not{\hbox{\kern-2pt $\del$}}}

\def\msb{{\bar{\ssstyle M \kern -1pt S}}}

\usepackage{color}
\usepackage{lineno}
\usepackage{subfig}
\usepackage{hyperref}
\usepackage{paralist}

\newcommand\pubnumber{PROC-CTD2020-12}

\newcommand\pubdate{\today}

\def\affiliation{
On behalf of the ALICE Collaboration, \\
CERN, Switzerland}

\newcommand{\conference}{Connecting the Dots Workshop (CTD 2020)\\
April 20-30, 2020}

\usepackage{fancyhdr}
\definecolor{mygrey}{RGB}{105,105,105}
\begin{document}

% uncomment the following line for adding line numbers
% \linenumbers

% large size for the first page
\large
\begin{titlepage}
\pubblock

%% Change the title, name, abstract
%% Title 
\vfill
\Title{Overview of online and offline reconstruction in ALICE for LHC Run 3}
\vfill

%  if you need to add the support use this, fill the \support definition above. 
%  \Author{FIRSTNAME LASTNAME \support}
\Author{David Rohr}
\Address{\affiliation}
\vfill

\begin{Abstract}
In LHC Run 3, ALICE will increase the data taking rate significantly to 50 kHz continuous readout of minimum bias Pb--Pb collisions.
The reconstruction strategy of the online-offline computing upgrade foresees a first synchronous online reconstruction stage during data taking enabling detector calibration, and a posterior calibrated asynchronous reconstruction stage.
The main challenges include processing and compression of 50 times more events per second than in Run 2, identification of removable TPC tracks and hits not used for physics, tracking of TPC data in continuous readout, the TPC space-charge distortion calibrations, and in general running more reconstruction steps online compared to Run 2.
ALICE will leverage GPUs to facilitate the synchronous processing with the available resources.
For the best GPU resource utilization, we plan to offload also several steps of the asynchronous reconstruction to the GPU.
In order to be vendor independent, we support CUDA, OpenCL, and HIP, and we maintain a common C++ source code that also runs on the CPU.
We will give an overview of the global reconstruction and tracking strategy, a comparison of the performance on CPU and different GPU models.
We will discuss the scaling of the reconstruction with the input data size, as well as estimates of the required resources in terms of memory and processing power.
\end{Abstract}

\vfill

% DO NOT CHANGE!!!
\begin{Presented}
\conference
\end{Presented}
\vfill
\end{titlepage}
\def\thefootnote{\fnsymbol{footnote}}
\setcounter{footnote}{0}
%

% normal size for the rest
\normalsize 

\section{Introduction}

ALICE (A Large Ion Collider Experiment) \cite{bib:alice} is one of the four major experiments at the LHC (Large Hadron Collider) at CERN.
It is a dedicated heavy-ion experiment studying lead collisions at the LHC at unprecedented energies.
After the second long LHC shutdown from 2019 to 2021, the LHC upgrade will provide a higher Pb--Pb collision rate, and ALICE will have updated many of its detectors and systems~\cite{bib:aliceupgrade}.
In particular, the main tracking detectors TPC (Time Projection Chamber) and ITS (Inner Tracking System) will be upgraded~\cite{bib:tpcrun3tdr}.
The trigger driven readout of Run 1 and Run 2 of up to 1\,kHz of Pb--Pb events is replaced by a continuous readout of all Pb--Pb events at 50\,kHz.
The continuous readout of pp collisions will happen at collision rates between~200\,kHz and~1\,MHz.
ALICE is abandoning the hardware triggers and will switch to a full online processing in software.
Also the computing scheme changes with the O$^2$ online-offline computing upgrade~\cite{bib:o2tdr}.
During data taking, the synchronous processing will serve two main objectives: detector calibration and data compression.
With a flat budget and the yearly increases of storage capacity, recording and storing raw data as today is prohibitively expensive at 50 times the data rate.
ALICE aims at a compression of the TPC data, the largest contributor to raw data size, of a factor 20 compared to the zero-suppressed raw data size of Run 2.
By producing the calibration during data taking, ALICE will reduce the number of offline reconstruction passes over the data, where the first two passes serve the calibration today.
The output of the synchronous data processing will be Compressed Time Frames (CTF).
The CTFs together with the calibration data will be stored to an on-site disk buffer, and from there replicated to tapes.
When the O2 computing farm is not fully used for the synchronous processing, e.g. in periods without beam or during pp data taking, it will perform a part of the asynchronous reconstruction, which reprocesses the data and generates the final reconstruction output.
The part of asynchronous processing that exceeds the capacity of the farm will be done in the Grid.
This asynchronous stage will employ the same algorithms and software as the synchronous stage, but with different settings, additional reconstruction steps, and final calibration.

\section{ALICE detector upgrades and implications for reconstruction}

Many detectors of ALICE are upgraded or replaced to operate at higher rates or with higher resolution.
The important changes with respect to track reconstruction are in particular the new Inner Tracking System (ITS) and the upgraded Time Projection Chamber (TPC).
The former ITS consisting of 2 layers of silicon pixels, 2 layers of silicon strips, and 2 layers of silicon drift detectors is replaced by 7 layers of silicon pixels improving the tracking close to the interaction region.
A new tracking algorithm based on the Cellular Automaton was developed for the new ITS that can optionally use GPUs~\cite{bib:itsgpu}.
For the TPC, the entire readout system is replaced.
During Run 1 and Run 2, the TPC was equipped with Multi Wire Proportional Chambers (MWPC) for the amplification and a gating grid to suppress the ion back flow.
The gating grid could operate at a maximum rate of around 3 kHz, which is incompatible to the continuous readout at 50 kHz.
Therefore, the MWPCs are replaced by Gas Electron Multipliers (GEM), which offer an intrinsic ion back flow blocking.
This allows for the removal of the gating grid and enables continuous readout.

Still, the ion back flow is significantly larger as during Run 2.
In particular, the majority of the ions are produced during the amplification, which means the ion density and the space-charge in the TPC scales linearly with the track occupancy and with the gain.
Therefore, the increase of the Pb--Pb interaction from 10 to 50 kHz adds another factor of 5 to the space-charge.
The electric field produced by the space-charge distorts the electrons during the drift leading in the worst case to a distortion of up to 20 cm for electrons created close to the central electrode at the inner radius of the TPC.
These distortions must be corrected down the the intrinsic TPC resolution that is in the order of 100\,$\mu$m.
This will require a new calibration procedure and has implications for the track reconstruction.

Also the change from trigger driven readout to continuous readout has implications for the track reconstruction.
For the triggered events, the absolute $z$ coordinate of TPC hits can be computed from the drift velocity, the time the hit was measured in the TPC, and the time of the triggered vertex.
The chicken and egg problem of the continuous readout is that the time of the vertex of a track is not known in advance but only after the hit is attached to a track and the track is attached to a vertex.
But without the vertex time the TPC hits cannot be converted from native pad, row, and time coordinates into $x$, $y$, and $z$ coordinates ahead of time.

The third important change is the switch from processing individual events to processing time frames containing 10 to 20 ms of continuous data.
In order to avoid complicating the reconstruction algorithms, all time frames are reconstructed independently.
TPC collisions happening at the end of a time frame have parts of their tracks in drift detectors like the TPC detected in the following time frame.
By construction, such tracks cannot be reconstructed.
This renders the last 90 $\mu$s (maximum drift time in the TPC) of a time frame unreconstructible.
A minimum time frame length of 10 ms is chosen such that the corresponding loss of statistics is below 1\%.

ALICE will neither employ hardware nor software triggers but store all data in compressed form.
This compression must happen in real time and process and store 50 times more collisions than during Run 2.

In summary, The new detectors and the new readout scheme  yield many challenges for data processing. The most imporant ones are:
\begin{itemize}
 \item Data must be strongly compressed in order to store 50 times as many compressed events as before.
 \item A calibration procedure for the space-charge distortions is required.
 \item Tracking in the TPC must work without absolute $z$ coordinates of TPC hits in continuous readout.
 \item Online processing speed must improve to processes 50 times more data than before.
 \end{itemize}
 
 \section{Data processing in Run 3}
 
 The first stage of online processing happens in the First Level Processors (FLP), a farm of around 200 compute nodes receiving the optical links from the detectors.
 The FLP performs mostly simple, local compression steps on the single link level.
 In addition, calibration and quality assurance tasks that require all data coming in from the same link must run here, since afterwards the data will be distributed among the compute nodes.
 The FLP nodes are equipped with up to 3 FPGA boards, which provide the optical transceivers and have the possibility to run a user logic for data processing.
 In the TPC case, for example, the FPGA performs the common mode correction and the zero suppression.
 Overall, the FLP farm reduces the data rate of 3.5 TB/s coming from the detectors down to 635 GB/s.
 
 The next stage is a farm of 250 to 500 Event Processing Nodes (EPN) housing in total around 2000 GPUs.
 The number of nodes will be decided when the final hardware is selected, which will define how many GPUs and how many processor cores are in one server.
 The EPN performs the bulk of the synchronous processing.
 Its data compression further reduces the incoming data rate from 635 GB/s to around 100 GB/s of CTFs.
 The CTFs are written to an on-site disk buffer and replicated to tape and to tier 0 and tier 1 data centers from there.
 
 While there is no beam in the LHC, or when the EPN farm is only partially used for pp data taking, the EPNs will contribute to the asynchronous reconstruction, reading the CTFs from the disk buffer and creating the final reconstruction output.
 The EPN contribution will be around one third of the total asynchronous reconstruction workload while the rest will be performed by the tier 0 and tier 1 data centers.
 
\section{TPC data compression}

The TPC is by far the largest contributor to the raw data rate among the detectors, with 3.4 TB/s.
Therefore, compression of the TPC data is most important and several elaborate compression steps are employed.
The projected compressed TPC data rate is 80 GB/s.
On the 10 GB/s level, other detectors contribute as well, so compression is applied but not as extensively as for the TPC.

The TPC compression runs in several steps:
\begin{compactenum}
 \item The raw data are clusterized and only the obtained hits are stored, which allows for a better entropy compression later on.
 \item All hit properties are stored in tailored custom fixed or floating point data formats using only exactly as many bits as needed to maintain the intrinsic TPC resolution.
 \item Hits of tracks not used for physics are removed. This includes tracks of very low momentum below $50$ MeV/$c$, secondary legs of looping tracks, track segments with high inclination angle that cannot be used in the track fit, and fake hits from noisy pads as well as charge clouds created by low momentum protons. Two strategies exist for the hit removal:
 \begin{compactenum}
   \item Strategy A foresees positive identification of all hits that can safely be removed. In addition, the online tracking identifies the good tracks used for physics and protects all hits in their vicinity from removal.
   \item Strategy B runs only the online tracking and removes all hits not in the vicinity of good tracks. By design, this includes all hits removed by strategy A. Therefore, strategy B yields the higher data reduction factor, and the shorter processing time. But the caveat is that tracks not found during the online tracking are inevitably lost. 
  \end{compactenum}
  Strategy A is the baseline solution, but both strategies are implemented and can be switched easily later after a careful evaluation.
  \item The entropy of the hit properties is reduced.
  \begin{compactitem}
   \item Hits attached to tracks are stored as residuals to the extrapolated track position, which have a smaller entropy than absolute hit positions.
   \item Hits not attached to tracks are sorted by geometrical coordinates and the difference to the previous hit is stored.
   \item Related hit properties, like cluster width, in different dimensions are treated together.
  \end{compactitem}
  \item The final step is entropy encoding. While Huffman compression was used during Run 2, ANS (Asymmetric Numeral Systems) \cite{bib:ans} encoding will be used in Run 3 to yield a higher compression.
\end{compactenum}
By design, the TPC compression will require full online tracking.
See \cite{bib:ctd2019} for more details.

\section{TPC calibration}

While there are many calibrations tasks, we focus on the TPC space-charge distortion (SCD) calibration, which is the most significant change compared to Run 2 and also the computationally most demanding task.
The SCD calibration will employ two methods described in the following sub-sections:

\subsection{Track based calibration}

This method was developed and employed during Run 2 to correct for the space-charge distortions, although with a much smaller magnitude than expected in Run 3.
In the track based calibration, TPC tracks are reconstructed uncalibrated with relaxed constraints and the tracks are matched to the inner detector (ITS) and the outer Transition Radiation Detector (TRD) and Time Of Flight (TOF) detector.
Although originally designed primarily for particle identification, these detectors contribute significant tracking information necessary for the TPC calibration.
This procedure  is possible since the distortions are smooth, such that the track seeding still works.
After the matching, the global track is refitted using only the information from the inner and outer detectors ignoring the attached TPC hits.
Afterwards, the residuals of the TPC hits with respect to the refitted track give a map of the distortions in the TPC.
The TPC volume is split in voxels, and inside each voxel the obtained distortions are inverted to obtain the correction map.
Postprocessing steps such as smoothing are applied to correct for holes in the acceptance of TRD and TOF and other effects.

The advantage of this method is that it automatically applies a couple of other calibration, e.g. drift velocity or misalignment.
However, a relatively large amount of tracks is needed to obtain the correction at the required precision, in particular since isolated tracks of peripheral collisions work best.
Due to the larger data sample, the Run 2 calibration intervals of 40 minutes will be shortened to about 1 minute in Run 3, but this is still insufficient to correct for short term fluctuations in the distortions.

One cause of changes in the distortion magnitude are changes in the LHC luminosity leading to changes in the TPC occupancy and thus in the space-charge.
This is compensated by scaling the correction map with the LHC luminosity.

But there are other causes of short term fluctuations like the LHC bunch structure and the Poissonian fluctuations of the number of collisions in a given time interval and collision centrality variations, which cannot be taken into account with the track based model.
The corrections for these effects must be on a much shorter time scale of around 5 ms.
These effects also existed during Run 2 but their order of magnitude was below the intrinsic TPC resolution.
This changes in Run 3 and the track based model alone will be insufficient for the SCD calibration.

\subsection{Integrated digital currents}

Another method is the integration of the currents arriving at the TPC pads by just accumulating the ADC values received on a channel.
From this one can compute the number of ions produced in the GEM amplification, the position of the ions in the TPC over time, the resulting space-charge, and the final distortions on the electrons.
These numerical computations are quite compute-intense and we aim at using a neural network to approximate the transformation from the currents to the correction map.
Compared to the track based calibration the only input is the number of ions, so this method cannot correct for any other effect but the SCD by design.
However, the integrated currents are available at a fine granularity such that short term corrections can be corrected for.

Naturally, this method requires the full ion history for a full ion drift time which is much longer than the electron drift time.
Therefore, it was impossible to apply this with the trigger driven readout in Run 2 and unfortunately the non-continuous Run 2 raw data cannot be used to evaluate this method.

\subsection{Combined method}

The full correction of the distortions requires the combination of both methods.
The track based calibration will correct for the average distortions.
The current based method will correct for the distortion fluctuations.
For this reason, the current based method is not applied on the currents themselves, but on the difference of the current and the average current.
In this way, we obtain the difference for the fluctuation correction which we add on the average correction obtained from the track based method.

The track based method will require online TPC, ITS, TRD, and TOF tracking in the EPN farm, but only for around 1\% of the total tracks selected from peripheral collisions.
The hit to track residuals will be extracted online, and a postprocessing step between the synchronous and the asynchronous reconstruction phases will produce the final correction maps.
The current based method will aggregate the currents already in the FPGA in the FLP farm.

\section{Track reconstruction with continuous data}

The processing of continuous data in the presence of space-charge distortions poses significant complications for the TPC tracking.
Position-dependent corrections like for SCD cannot be applied ahead of time for TPC hits.
The same holds for other effects like inhomogeneities in the magnetic field or the hit error parameterization.
Therefore, the TPC tracking runs first standalone without such corrections.
Since the effects are smooth, the track finding is not heavily affected.
Track seeds are extrapolated to the beam line and the most likely $z$ coordinate is computed under the assumption that the track is a primary and the vertex is at the origin.
The track is refit with the corresponding corrections and then matched to ITS tracks, which are reconstructed standalone in the ITS.
This matching fixes the time of the TPC track and enables the propagation into the outer detectors.
The tracking algorithm has been derived from the Run 2 High Level Trigger (HLT)~\cite{bib:hltpaper} and modified to suit the needs of Run 3.
For more details see \cite{bib:ctd2018}

\section{Reconstruction chain and performance}

Summarizing the processing requirements, the synchronous reconstruction performs the calibration and the data compression.
The TPC data compression requires full TPC tracking, both for the rejection of hits and for the track model compression.
Other detectors perform only the ANS entropy compression and optionally clusterization.
The calibration requires tracking for the other detectors as well, but only for a small fraction of events in the order of 1\%.
Therefore, the dominant part of synchronous processing is the TPC tracking.
ALICE will employ GPUs to speed up the TPC processing as it did during Run 1 and Run 2 in the HLT~\cite{bib:hltpaper}.
The TPC processing time on the GPU defines the number of GPUs required in the EPN farm, and thus the size of the farm itself.

While TPC tracking is also a significant fraction of the asynchronous reconstruction, it is not so dominant.
First, the synchronous reconstruction already removed a significant fraction of the hits, which in turn speeds up the asynchronous reconstruction.
Second, several tracking steps like the following of looping tracks are not needed in the asynchronous reconstruction.
And last, all other detectors have to process all events in the asynchronous reconstruction compared to O(1\%) in the synchronous.
This makes for instance ITS tracking also computationally intensive.
Without using the GPUs in additional steps of the asynchronous reconstruction, they would be idling most of time time, considering that synchronous data taking of Pb--Pb events happens only during a few weeks in a year.
ALICE is therefore aiming to use the GPUs also in many places during the asynchronous reconstruction, and a promising case is the full tracking chain of the central barrel region.

\begin{figure}[!htb]
  \centering
  \includegraphics[width=0.99\textwidth]{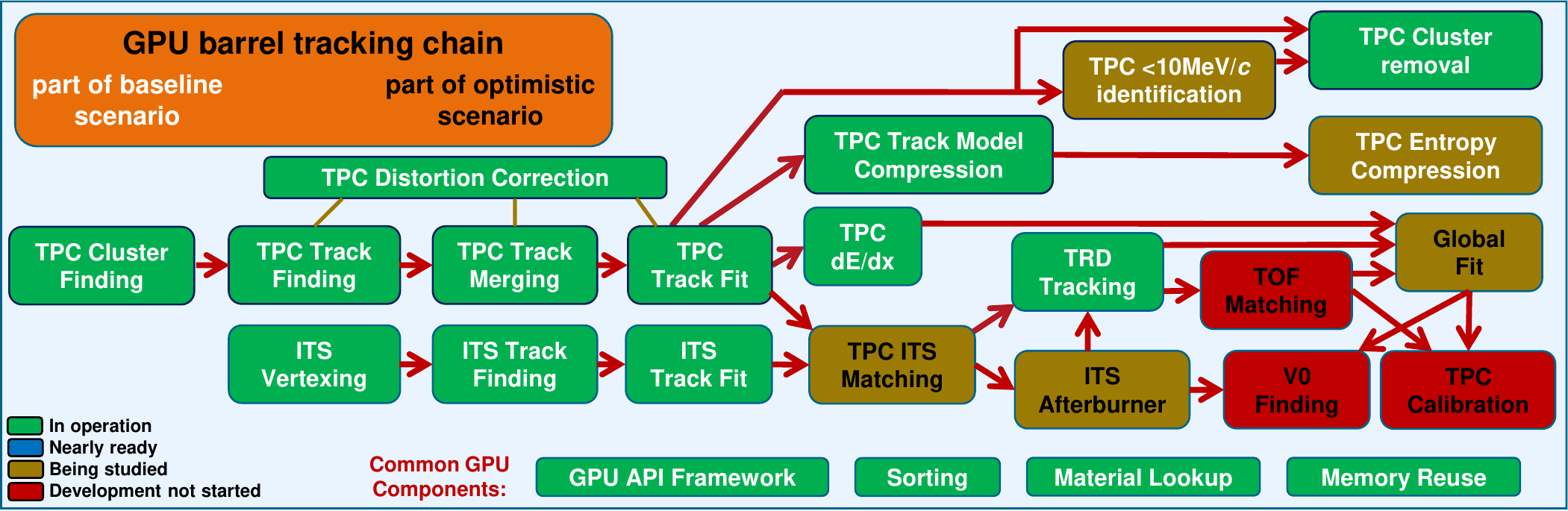}
  \caption{Overview of the compute-intense reconstruction steps of the global barrel tracking chain and their state of GPU usage.}
  \label{fig:components}
\end{figure}

Figure~\ref{fig:components} gives an overview of all barrel tracking steps that are promising candidates for the GPU in the long run.
The baseline scenario is what is needed during the synchronous reconstruction to keep up with 50 kHz Pb--Pb data, and the required processing steps have been almost fully implemented on the GPU.
Some consolidation and testing will be needed.
Afterwards the focus will shift on the optimistic scenario to eventually port many if not all of these steps onto the GPU for the asynchronous reconstruction.

\subsection{Processing large time frames}

Several HEP experiments recording mostly pp events are struggling to utilize GPUs to the full extent since single collisions do not exhibit sufficient parallelism.
General approaches are to combine many events into chunks and process them on the GPU in one go.
While a heavy ion collision could fully load a GPU at the time of LHC Run 1, this is no longer the case today for modern GPUs with many more compute units.
However, the time frames recorded by ALICE in Run 3 contain hundreds of such collisions, which will be sufficient parallelism also for future GPUs.
Instead, the memory becomes the limitation since the full time frame and all temporary memory for its processing must fit inside the GPU.
Therefore, memory usage is being optimized as much as possible.
Processing of triggered events can happen on the GPU independenly, which enables  the reusage of GPU memory used for one collision as soon as it is finished.
In the ALICE case, the full time frame with hundreds of collisions is processed at once and must not be split.
Therefore, the memory is not reused for independent collisions but for consecutive reconstruction steps.
More details are given in~\cite{bib:chep2019}.
Overall, this poses a limit on the time frame size, and ALICE is currently working with a compromise of 10 ms as the default length.
This should eventually require around 16 GB of GPU memory and the unreconstructible ends of time frames will correspond to less than 1\% of the total data.

\begin{figure}[!htb]
  \centering
  \subfloat[]{\includegraphics[width=0.47\linewidth]{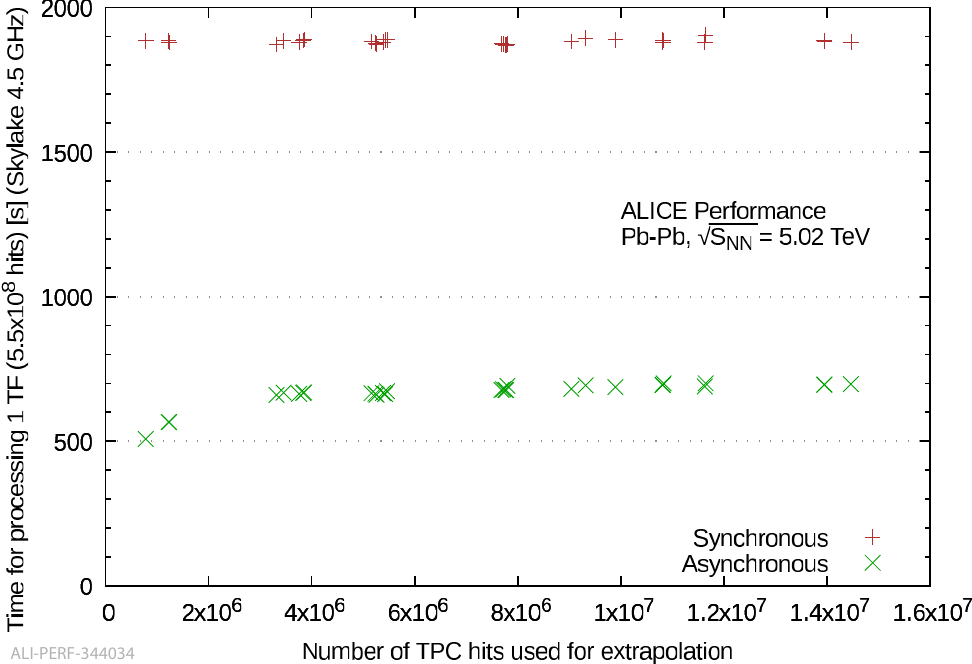}}
  \qquad
  \subfloat[]{\includegraphics[width=0.47\linewidth]{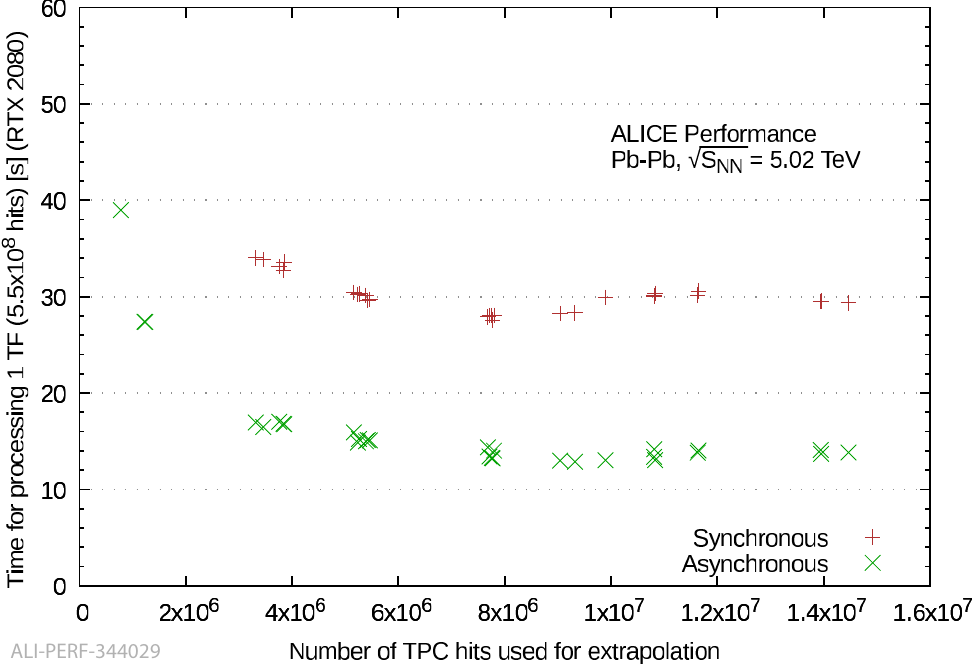}}
  \caption{Estimated processing time for one full time frame of 256 LHC orbits ($\sim20$ ms). The time is extrapolated linearly in the number of hits from the measurements of the processing time of shorter time frames. Times are measured on CPU (a) and GPU (b).}
  \label{fig:perf}
\end{figure}

It is important to understand how the processing time is affected by the length of the time frame, in particular since the memory optimization is still ongoing and we still cannot process a full time frame on the GPUs we have.
If, for instance, the processing time would grow more than linearly with the input data size, it would be imperative from the computing perspective to use shorter time frames or to optimize the algorithm.
Figure~\ref{fig:perf} shows the extrapolated time (linearly in the number of TPC hits) for processing a full time frame based on the processing time of shorter time frames.
The y axis shows the extrapolated processing time, the x axis shows the number of clusters.
The distribution is pretty much flat for the CPU (Fig.~\ref{fig:perf} (a)), with only a slight speed-up for the smallest measurement on the left due to cache effects.
In case of the GPU (Fig.~\ref{fig:perf} (b)), the smaller time frames take relatively longer for processing because the parallelism is limited, but starting from 8 million hits in the TPC, the processing time per hit remains more or less constant.

\subsection{Comparison of GPUs and CPUs}

Table~\ref{tab:perf} gives an overview of the processing times of several reconstruction steps on different GPU models and on an Intel CPU using a single core.
For reference, we have measured the multi-core scalability of the CPU processing many independent events in parallel.
Using 64 threads on an AMD Rome 64 core CPUs, we achieved a speed-up of around 54x of the overall processing time on the CPU.
Using the SMT feature and all 128 threads, we see a speedup of around 70x.

\begin{table}[!htb]
  \begin{center}
    \begin{tabular}{l|rrrr}
      \hline
      \hline
      Processing step & AMD Radeon 7 & RTX 2080 Ti & Intel CPU & 2080 Ti / CPU \\
      \hline
      Zero Suppression Decoding & 38 ms & 19 ms & 986 ms & 52x \\
      Cluster Finding & 87 ms & 79 ms & 21343 ms & 270x \\
      Track Finding & 109 ms & 65 ms & 8759 ms & 135x \\
      Track Fit & 284 ms & 243 ms & 7204 ms & 30x \\
      Cluster Compression & 137 ms & 105 ms & 1452 ms & 14x \\
      Synchronous Processing Total & 657 ms & 511 ms & 39816 ms & 78x \\
      d$E$/d$x$ Calculation & 61 ms & 22 ms & 906 ms & 41x \\
      Asynchronous Processing Total & 304 ms & 237 ms & 13381 ns & 56x\\
      \hline
      \hline
    \end{tabular}
    \caption{Processing time of reconstruction steps on GPU and CPU. The CPU was measured on an Intel CPU clocked at 4.5 GHz with the Skylake architecture (clock fixed, turbo disabled) on a single core.}
    \label{tab:perf}
  \end{center}
\end{table}

\begin{table}[!htb]
  \begin{center}
    \begin{tabular}{l|rr}
      \hline
      \hline
      GPU Model & Performance (relative) & Memory \\
      \hline
      NVIDIA RTX 2080 Ti & 100.0\% & 11 GB \\
      NVIDIA Quadro RTX 6000 (active cooling) & 105.8\% & 24 GB \\
      NVIDIA Quadro RTX 6000 (passive cooling) & 94.5\% & 24 GB \\
      NVIDIA RTX 2080 & 83.5\% & 8 GB \\
      NVIDIA GTX 1080 & 60.1\% & 8 GB \\
      NVIDIA V100 & 88.5\% & 32 GB \\
      NVIDIA T4 & 59.3\% & 16 GB \\
      AMD Radeon 7 & 77.8\% & 16 GB \\
      AMD MI50 & 74.1\% & 16 GB \\
      Intel Skylake 4.5 GHz (1 core) & 1.77\% & - \\
      \hline
      \hline
    \end{tabular}
    \caption{Relative performance in the synchronous processing stage normalized to the performance of an NVIDIA RTX 2080 Ti. In case of the NVIDIA GTX 1080, all kernels that use thrust sort were excluded from the comparison, due to a severe performance degradation with that architecture. In the case of the Intel CPU, only the steps used in the asynchronous phase are compared, since the cluster finding of the synchronous phase is implemented in a way that is very slow on the CPU and thus not suited for the comparison.}
    \label{tab:perfrel}
  \end{center}
\end{table}

Table~\ref{tab:perfrel} shows the relative processing time on several GPU models.
Several particularities should be noted.
Some steps (in Table~\ref{tab:perf} the d$E$/d$x$ and the Track Fit) make extensive use of sort algorithms.
We observed a severe performance degradation in these sorting steps on the GTX 1080 GPU, which we didn't investigate further since this GPU is end of life and not a candidate for our farm.
Therefore, we excluded these kernels from the measurement of the GTX 1080.
Table~\ref{tab:perf} shows that the cluster finding is extremely slow on the processor because the algorithm was developed for GPUs in a way that is very sub optimal for the CPU.
The clusterization happens only during the synchronous processing which will anyway happen on the GPU, therefore its CPU performance is mostly irrelevant.
For a fair comparison, we thus exclude the clusterization from the relative CPU performance value.
Overall, an NVIDIA RTX 2080 Ti can replace around 56 CPU cores.
This is in agreement with previous measurements, e.g. in~\cite{bib:hltpaper} we reported that a GTX 1080 can replace around 40 CPU cores at the end of Run 2.
Considering that the number of CPU cores is increasing and we are comparing to a single core, the ratio of GPU to CPU performance (comparing all shaders of a GPU to all cores of a CPU) didn't change much with respect to our previous measurements~\cite{bib:hltpaper}.
In particular, the new GPUs have more compute units as well.

The most promising GPU candidates for the EPN farm are the AMD Radeon 7, AMD MI 50, NVIDIA RTX 2080 Ti, and NVIDIA Quadro RTX 6000.
With all GPUs, it is possible to build the EPN farm using around 2000 cards, in the NVIDIA case with even less.
The final decision will be taken based on the costs of the overall farm.
We did not observe large performance differences between the consumer grade GPUs (RTX 2080 Ti, Radeon 7) and the professional GPU (Quadro RTX, MI50).
An important difference for us is the possibility for passive cooling in the server and continuous support.
The NVIDIA RTX 2080 Ti with only 11 GB of memory would obviously require an additional shortening of the time frame below 10 ms, in case of the two AMD GPUs it is yet not fully clear whether the 16 GB will be sufficient for a 10 ms time frame, while the Quadro RTX 6000 will certainly be able to process the full 10 ms time frame.

\section{Conclusions}

For LHC Run 3, ALICE will switch to continuous readout of 50 kHz minimum bias Pb--Pb events without trigger.
ALICE will collect 50 times as many events as before.
Full online processing will happen in software mostly on GPUs.
During data taking, the synchronous processing will perform data compression and calibration.
Computationally, the dominant part is the TPC tracking which will run on the GPUs.
When there is no beam, the asynchronous processing will produce the final reconstruction output, and we aim to have as many processing steps as possible on the GPU.
The most demanding processing tasks will be the TPC compression, the space-charge distortion calibration, and the tracking.
TPC compression will be based on the Run 2 strategy, and in addition we will employ track model compression, remove hits not used for physics, and switch from Huffman to ANS encoding.
The TPC SCD calibration will combine two methods: the track based method will calibrate for the average distortions and the current based method will correct for the fluctuations that come on top.
The TPC tracking is derived from the Run 2 HLT tracking and was improved to achieve the same resolution as the Run 2 offline tracking and adapted to process continuous data.
There are promising GPU candidates from both AMD and NVIDIA, which will enable ALICE to do full online processing with less than 2000 GPUs.

%%%%%%%%%%%%%%%%%%%%%%%%%%%%%%%%%%%%%%%%%%%%%%%%%%%%%%%%%%%%%%%%%%%%%%%%%%%

\end{document}